\def\fileversion{v2.5}
\def\filedate{4 October 1993}

\newdimen\@bls                    
\@bls=\baselineskip               
\advance\@bls -1ex                
\newdimen\@eps                    %
\def\section{\@startsection{section}{1}{\z@}
  {1.5\@bls plus 0.5\@bls}{1\@bls}{\normalsize\bf}}
\def\subsection{\@startsection{subsection}{2}{\z@}
  {1\@bls plus 0.25\@bls}{\@eps}{\normalsize\bf}}
\def\subsubsection{\@startsection{subsubsection}{3}{\z@}
  {1\@bls plus 0.25\@bls}{\@eps}{\normalsize\bf}}
\def\paragraph{\@startsection{paragraph}{4}{\parindent}
  {1\@bls plus 0.25\@bls}{0.5em}{\normalsize\bf}}
\def\subparagraph{\@startsection{subparagraph}{4}{\parindent}
  {1\@bls plus 0.25\@bls}{0.5em}{\normalsize\bf}}

\def\@sect#1#2#3#4#5#6[#7]#8{\ifnum #2>\c@secnumdepth
  \def\@svsec{}\else
  \refstepcounter{#1}\edef\@svsec{\csname the#1\endcsname.\hskip0.5em}\fi
  \@tempskipa #5\relax
  \ifdim \@tempskipa>\z@
    \begingroup
      #6\relax
      \@hangfrom{\hskip #3\relax\@svsec}{\interlinepenalty \@M #8\par}%
    \endgroup
    \csname #1mark\endcsname{#7}\addcontentsline
      {toc}{#1}{\ifnum #2>\c@secnumdepth \else
        \protect\numberline{\csname the#1\endcsname}\fi #7}%
  \else
    \def\@svsechd{#6\hskip #3\@svsec #8\csname #1mark\endcsname
      {#7}\addcontentsline{toc}{#1}{\ifnum #2>\c@secnumdepth \else
        \protect\numberline{\csname the#1\endcsname}\fi #7}}%
  \fi \@xsect{#5}}

\long\def\@makefigurecaption#1#2{\vskip 10mm #1. #2\par}

\long\def\@maketablecaption#1#2{\hbox to \hsize{\parbox[t]{\hsize}
  {#1 \\ #2}}\vskip 0.3ex}

\def\fnum@figure{Figure \thefigure}
\def\figure{\let\@makecaption\@makefigurecaption \@float{figure}}
\@namedef{figure*}{\let\@makecaption\@makefigurecaption \@dblfloat{figure}}

\def\table{\let\@makecaption\@maketablecaption \@float{table}}
\@namedef{table*}{\let\@makecaption\@maketablecaption \@dblfloat{table}}

\textfloatsep=\floatsep           
\intextsep=\floatsep              

\long\def\@makefntext#1{\parindent 1em\noindent\hbox{${}^{\@thefnmark}$}#1}

\def\maketitle{\begingroup        
    \def\thefootnote{\fnsymbol{footnote}}%
    \newpage \global\@topnum\z@
    \@maketitle \@thanks
  \endgroup
  \let\maketitle\relax \let\@maketitle\relax
  \gdef\@thanks{}\let\thanks\relax
  \gdef\@address{}\gdef\@author{}\gdef\@title{}\let\address\relax}

\def\justify@on{\let\\=\@normalcr
  \leftskip\z@ \@rightskip\z@ \rightskip\@rightskip}

\newbox\fm@box                    

\def\@maketitle{
  \global\setbox\fm@box=\vbox\bgroup
    \vskip 8mm                    
    \raggedright                  
    \hyphenpenalty\@M             
    {\Large \@title \par}         
    \vskip\@bls                   
    {\normalsize                  
     \@author \par}               
    \vskip\@bls                   
    \@address                     
  \egroup
  \twocolumn[
    \unvbox\fm@box                
    \vskip\@bls                   
    \unvbox\abstract@box          
    \vskip 2pc]}                  

\newcounter{address}
\def\theaddress{\alph{address}}
\def\@makeadmark#1{\hbox{$^{\rm #1}$}}

\def\address#1{\addressmark\begingroup
  \xdef\@tempa{\theaddress}\let\\=\relax
  \def\protect{\noexpand\protect\noexpand}\xdef\@address{\@address
  \protect\addresstext{\@tempa}{#1}}\endgroup}
\def\@address{}

\def\addressmark{\stepcounter{address}%
  \xdef\@tempa{\theaddress}\@makeadmark{\@tempa}}

\def\addresstext#1#2{\leavevmode \begingroup
  \raggedright \hyphenpenalty\@M \@makeadmark{#1}#2\par \endgroup
  \vskip\@bls}

\newbox\abstract@box              

\def\abstract{%
  \global\setbox\abstract@box=\vbox\bgroup
  \small\rm
  \ignorespaces}
\def\endabstract{\par \egroup}

\def\thebibliography#1{\section*{REFERENCES}\list{\arabic{enumi}.}
  {\settowidth\labelwidth{#1.}\leftmargin=1.67em
   \labelsep\leftmargin \advance\labelsep-\labelwidth
   \itemsep\z@ \parsep\z@
   \usecounter{enumi}}\def\makelabel##1{\rlap{##1}\hss}%
   \def\newblock{\hskip 0.11em plus 0.33em minus -0.07em}
   \sloppy \clubpenalty=4000 \widowpenalty=4000 \sfcode`\.=1000\relax}

\newcount\@tempcntc
\def\@citex[#1]#2{\if@filesw\immediate\write\@auxout{\string\citation{#2}}\fi
  \@tempcnta\z@\@tempcntb\m@ne\def\@citea{}\@cite{\@for\@citeb:=#2\do
    {\@ifundefined
       {b@\@citeb}{\@citeo\@tempcntb\m@ne\@citea
        \def\@citea{,\penalty\@m\ }{\bf ?}\@warning
       {Citation `\@citeb' on page \thepage \space undefined}}%
    {\setbox\z@\hbox{\global\@tempcntc0\csname b@\@citeb\endcsname\relax}%
     \ifnum\@tempcntc=\z@ \@citeo\@tempcntb\m@ne
       \@citea\def\@citea{,\penalty\@m}
       \hbox{\csname b@\@citeb\endcsname}%
     \else
      \advance\@tempcntb\@ne
      \ifnum\@tempcntb=\@tempcntc
      \else\advance\@tempcntb\m@ne\@citeo
      \@tempcnta\@tempcntc\@tempcntb\@tempcntc\fi\fi}}\@citeo}{#1}}

\def\@citeo{\ifnum\@tempcnta>\@tempcntb\else\@citea
  \def\@citea{,\penalty\@m}%
  \ifnum\@tempcnta=\@tempcntb\the\@tempcnta\else
   {\advance\@tempcnta\@ne\ifnum\@tempcnta=\@tempcntb \else \def\@citea{--}\fi
    \advance\@tempcnta\m@ne\the\@tempcnta\@citea\the\@tempcntb}\fi\fi}

\def\ps@crcplain{\let\@mkboth\@gobbletwo
     \def\@oddhead{\reset@font{\sl\rightmark}\hfil \rm\thepage}%
     \def\@evenhead{\reset@font\rm \thepage\hfil\sl\leftmark}%
     \let\@oddfoot\@empty
     \let\@evenfoot\@oddfoot}

\ps@crcplain                    

\documentstyle[twoside,fleqn,espcrc2]{article}
\newcommand{\be}{\begin{equation}}
\newcommand{\ee}{\end{equation}}
\newcommand{\bea}{\begin{eqnarray}}
\newcommand{\eea}{\end{eqnarray}}
\newcommand{\bd}{\begin{displaymath}}
\newcommand{\ed}{\end{displaymath}}

\newcommand{\AmS}{{\protect\the\textfont2
  A\kern-.1667em\lower.5ex\hbox{M}\kern-.125emS}}

\title{ Topological features in a two-dimensional Higgs model }

\author{I. Montvay\address{Deutsches Elektronen-Synchrotron DESY, \\
        Notkestr. 85, D-22603 Hamburg, FRG}}

\begin{document}

\begin{abstract}
 Topological properties of the gauge field in a two-dimensional Higgs
 model are investigated.
 Results of exploratory numerical simulations are presented.
\end{abstract}

\maketitle

\section{ INTRODUCTION }
 Fermion number, which is the sum of baryon number and lepton number
 ($B+L$), is not conserved in the Standard Model \cite{THOOFT}.
 This is due to the anomaly in the fermion current.

 The lattice formulation of the anomalous fermi\-on number
 non-conservation is problematic \cite{BANKS}, because it has to do with
 chiral gauge couplings and, as is well known, there is a difficulty
 with chiral gauge theories on the lattice (see, for instance, the
 reviews \cite{TSUKPR,PETCHER}).
 There is, however, an approximation of the electroweak sector of the
 standard model which can be studied with standard lattice techniques,
 namely the limit when the
 $\rm SU(3)_{colour} \otimes U(1)_{hypercharge}$ gauge couplings are
 neglected \cite{AMSTPR,JOHOPR}.

 A simple prototype model is the standard $\rm SU(2)_L$ Higgs model
 coupled to an even number of fermion doublets.
 For a numerical study of the bounds on renormalized couplings in
 such a model without $\rm SU(2)_L$ gauge field see the contribution of
 Gernot M\"unster in this proceedings \cite{BOUNDS}.

 Before doing numerical simulations in the four dimensional
 $\rm SU(2)_L$ gauge model for anomalous fermion number violation, it
 is useful to study a simple U(1) toy model in two dimensions, which has
 often been studied in this context (see e.~g. \cite{BOCSHA}).
 The lattice formulation of an appropriate two-dimensional U(1) Higgs
 model has been summarized in \cite{STLOUIS}.

 The lattice action depending on the compact U(1) gauge field
 $U_{x\mu}=\exp(iA_\mu(x)),\; (\mu=1,2)$ and, for simplicity, fixed
 length Higgs scalar field $\phi(x),\; |\phi(x)|=1$ has two parameters:
 the inverse gauge coupling squared $\beta=1/g^2$  and the hopping
 parameter of the scalar field $\kappa$:
\bd
S = \beta \sum_x \sum_{\mu=1, \nu=2}
[1 - cos(F_{\mu\nu}(x))]
\ed
\be \label{eq01}
-2\kappa \sum_x \sum_{\mu=1}^2 \phi^*(x+\hat{\mu})U_{x\mu}\phi(x) \ .
\ee
 Here the lattice gauge field strength is defined for $\mu,\nu=1,2$ as
\bd
F_{\mu\nu}(x) =
\ed
\be \label{eq02}
A_\nu(x+\hat{\mu})-A_\nu(x)-A_\mu(x+\hat{\nu})+A_\mu(x) \ .
\ee
 Real angular variables $-\pi < \theta_{x\mu} \leq \pi$ on the links
 can be introduced by
\bd
U_{x\mu} \equiv \exp(i\theta_{x\mu}) \ ,
\ed
\be \label{eq03}
\theta_{x\mu} = A_\mu(x)-2\pi \cdot \mbox{\rm NINT}(A_\mu(x)/2\pi) \ ,
\ee
 where NINT() denotes nearest integer.

 Fermions in this two dimensional model are introduced in the
 mirror fermion basis $(\psi,\chi)$.
 The continuum limit of the anomaly equation for the two dimensional
 current, which corresponds to the fermion number current in the four
 dimensional $\rm SU(2)_L$ model, is the following \cite{STLOUIS}:
\be \label{eq04}
\langle \partial_\mu J_\mu(x) \rangle
= \frac{1}{2\pi} \epsilon_{\mu\nu} F_{\mu\nu}(x) = 2 q(x) \ .
\ee
 Here $q(x)$ is the density of the topological charge
\be \label{eq05}
q(x)= \frac{1}{4\pi} \epsilon_{\mu\nu} F_{\mu\nu}(x) \ .
\ee

 Since according to (\ref{eq04}) the non-conservation of the fermion
 current $J_{x\mu}$ is proportional to the topological charge
 density, the first step in understanding the fermion number anomaly on
 the lattice is to understand the topological features of the two
 dimensional U(1) lattice gauge fields.
\begin{figure}
\vspace{7cm}
\includegraphics{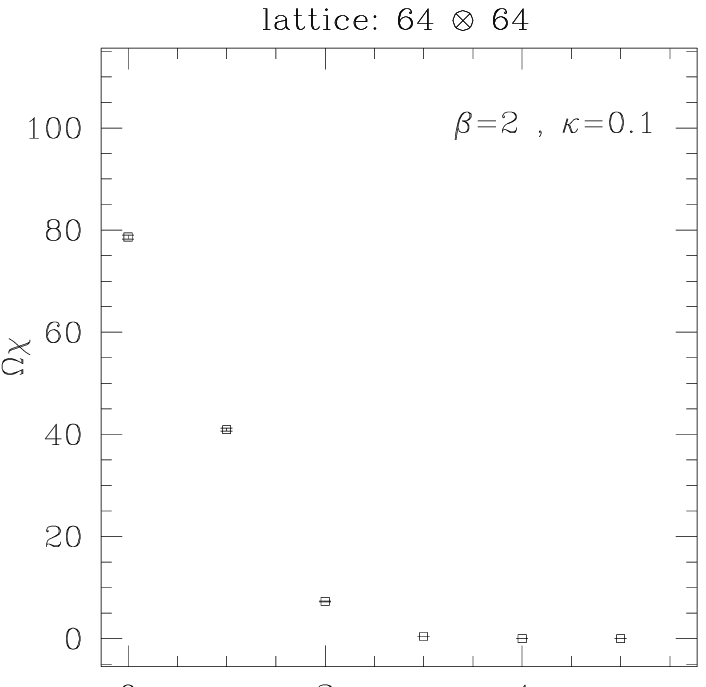}
\caption{ \label{figb2}
The topological susceptibility on block\-ed configurations
as a function of blocking level at $\beta=2.0,\;\kappa=0.1$
on $64^2$ lattice.
$\Omega\chi$ is plotted.}
\end{figure}

\section{ TOPOLOGICAL CHARGE }
 The topological charge of U(1) lattice gauge field configurations
 can be defined as a sum over the contributions of plaquettes.
 The basic assumption is the existence of a piecewise continuous
 interpolation of the gauge field \cite{LUSCH,GKSW}.
 The gauge invariant topological charge on the torus corresponding to
 periodic boundary conditions is obtained either from the
 ``transition functions''  or from the ``sections'' of this
 interpolated gauge field.

 Introducing the plaquette angles \\
 $-4\pi < \Theta_{x\mu\nu} \leq 4\pi$ and
 $-\pi < \theta_{x\mu\nu} \leq \pi$ by
\bd
\Theta_{x\mu\nu} \equiv
\theta_{x\mu} + \theta_{x+\hat{\mu},\nu} -
\theta_{x\nu} - \theta_{x+\hat{\nu},\mu} \ ,
\ed
\be \label{eq06}
\theta_{x\mu\nu} \equiv
\Theta_{x\mu\nu} - 2\pi \cdot \mbox{\rm NINT}(\Theta_{x\mu\nu}/2\pi) \ ,
\ee
 one can show that the total topological charge $Q$ is given by
\be \label{eq07}
Q = \frac{1}{2\pi} \sum_x \theta_{x12}
= - \sum_x \mbox{\rm NINT}(\Theta_{x12}/2\pi) \ .
\ee
 An interesting physical quantity is the topological susceptibility
\be \label{eq08}
\chi \equiv \Omega^{-1}(\langle Q^2 \rangle - \langle Q \rangle^2) \ ,
\ee
 where $\Omega \equiv L_1 \cdot L_2$ denotes the number of lattice
 points.
 (Note that in \cite{JOHOPR} $Q$ has been defined with opposite sign
 and $\chi$ without the normalization factor $\Omega^{-1}$.)
\begin{figure}
\vspace{7cm}
\includegraphics{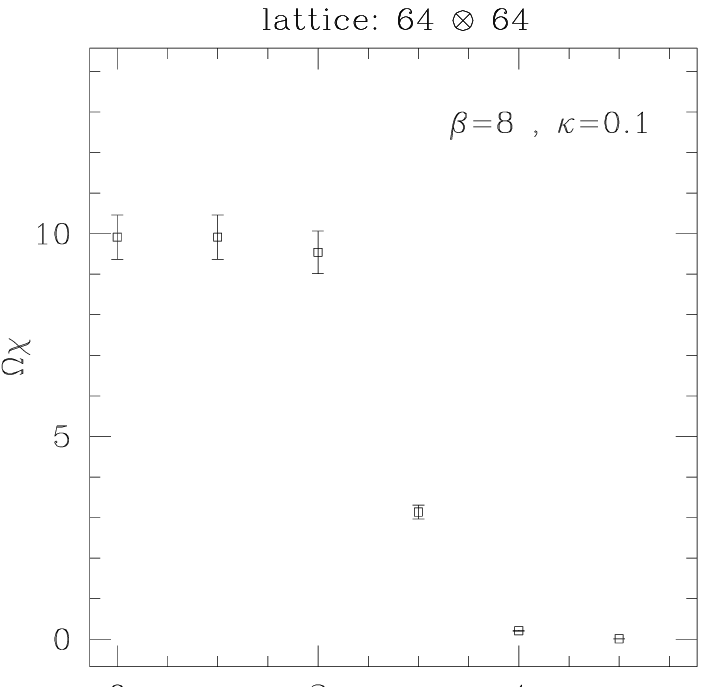}
\caption{ \label{figb8}
The topological susceptibility on block\-ed configurations
as a function of blocking level at $\beta=8.0,\;\kappa=0.1$
on $64^2$ lattice.
$\Omega\chi$ is plotted.}
\end{figure}

\subsection{Blocking and cooling}
\begin{figure}
\vspace{7cm}
\includegraphics{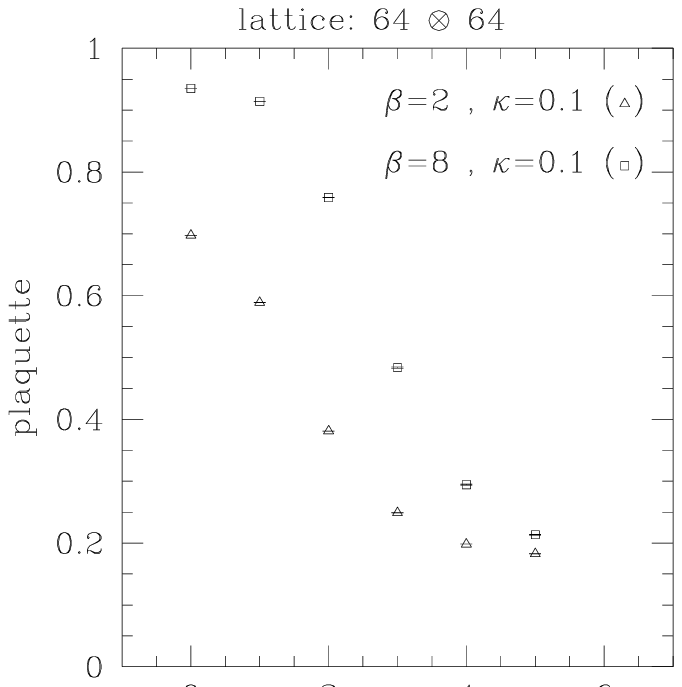}
\caption{ \label{figbp}
The average plaquette on blocked configurations as a function of
blocking level at $\beta=2.0,\;\kappa=0.1$ (triangles)
and $\beta=8.0,\;\kappa=0.1$ (squares).
The original lattice is $64^2$. }
\end{figure}

 The total topological charge is a long distance property of the gauge
 configuration, therefore it should remain stable under smoothing
 the short range fluctuations, for instance, by blocking.
 This indeed happens in a few blocking steps.
 Defining the $2^n$ blocked gauge fields in the usual way by averaging
 over the two ``staples'' and a straight line connecting next nearest
 neighbour sites, one obtains, for instance, figs. \ref{figb2},
 \ref{figb8}.
 As one can see, at $\beta=8.0$ the topological charge is kept in two
 blocking steps but at $\beta=2.0$ already one blocking step is
 enough to decrease it substantially.
 The important variable is the plaquette expectation value
 $P \equiv \langle \cos\Theta_{x12} \rangle$ (see fig. \ref{figbp}).
 Below $P \simeq 0.6$ the topological charge is erased by blocking.
 This means that for gauge configurations with smaller average
 plaquette value the lattice definition of the topological charge
 is problematic (impossible?).

 One can also investigate the effect of locally minimizing the
 gauge field action by {\em cooling} sweeps.
 It turns out that at larger gauge couplings, e.~g. at $\beta=2.0$,
 one can observe a slight decrease in the topological susceptibility but
 at $\beta=8.0$, after several tens of cooling sweeps, there is no
 change at all.

\subsection{Topological susceptibility}
\begin{figure}
\vspace{5.0cm}
\includegraphics{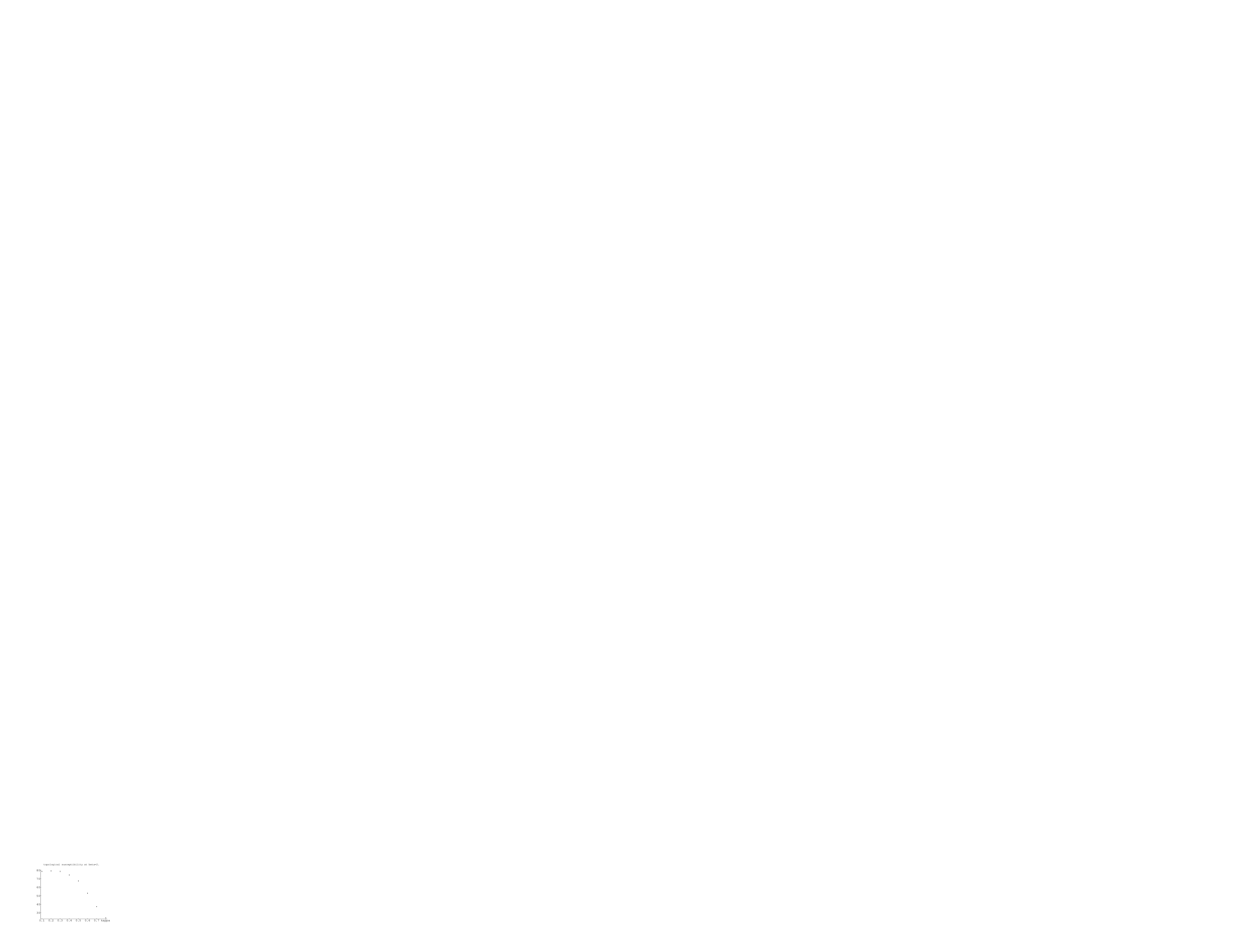}
\caption{ \label{figt2}
The topological susceptibility as function of $\kappa$
at $\beta=2.0$ on $64^2$ lattice.
$\Omega\chi$ is plotted.}
\end{figure}

\begin{figure}
\vspace{5.5cm}
\includegraphics{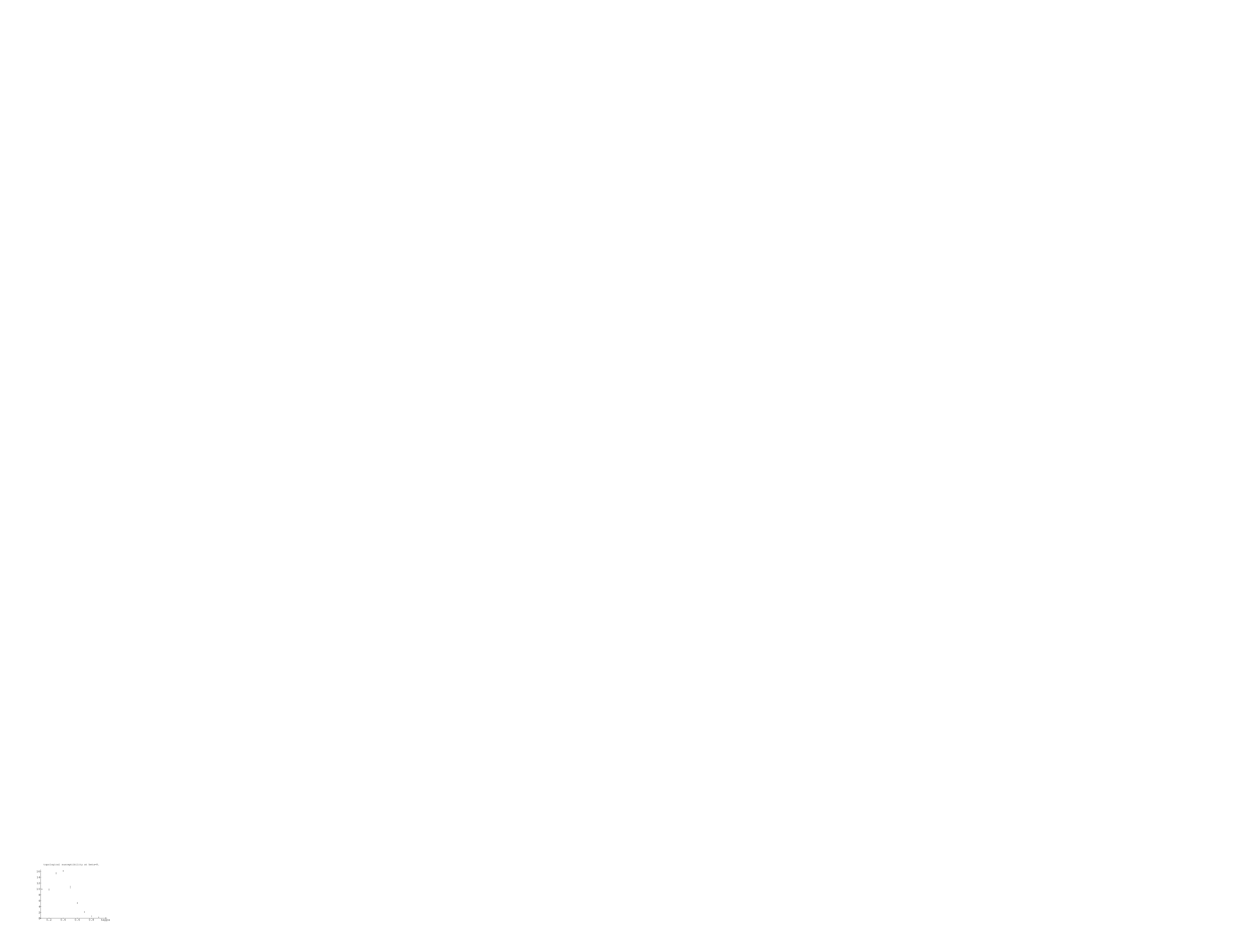}
\caption{ \label{figt8}
The topological susceptibility as function of $\kappa$
at $\beta=8.0$ on $64^2$ lattice.
$\Omega\chi$ is plotted.}
\end{figure}

 In a series of Monte Carlo runs the topological susceptibility has
 been measured on $64^2$ lattices at $\beta=2.0$ and $\beta=8.0$ as
 a function of the scalar hopping parameter $\kappa$ (see figs.
 \ref{figt2}, \ref{figt8}).
 At every point 60000 to 90000 measurements were collected, separated
 by one Metropolis and 12 overrelaxation updating steps.
 The expectation value of $Q$ turns out to be small, as it should.
 For instance, at $\beta=8.0,\; \kappa=0.2$ one gets
 $\langle Q \rangle=-0.5(6)$.
 As one can see, $\chi$ is considerably smaller at $\beta=8.0$, and
 at this small gauge coupling there is a maximum near $\kappa=0.4$,
 which presumably corresponds to the continuation of the
 Kosterlitz-Thouless transition at $\beta=\infty$ for finite $\beta$.

\section{ GAUGE FIXING AND MODIFIED ACTION }
 A plaquette contributing by a non-zero integer to the right hand
 side of (\ref{eq07}) can be imagined to carry a {\em ``Dirac string''}
 or {\em ``gauge field kink''} \cite{DEGTOU}.
 These local contributions to the topological charge play an
 important r\^ole in the dynamics of lattice U(1) gauge fields.
 However, it is important to keep in mind that the individual terms on
 the right hand side of (\ref{eq07}) are not gauge invariant, in
 contrast to the total topological charge $Q$ (and the terms in the
 first form of $Q$ in (\ref{eq07})).
 Namely, performing large gauge transformations on the two ends of a
 link can create or annihilate a kink-antikink pair on the two
 plaquettes which both contain the link.

 In order to give a physical meaning to the number of gauge kinks one
 has to fix the gauge.
 Since the plaquettes with $\Theta_{12} \simeq \pm 2\pi$ and
 $|\theta_\mu| \simeq \pi/2$ contain large link angles,
 one possibility is to minimize the sum of squares of the link angles:
 $ \sum_x \sum_{\mu=1}^2 \theta_{x\mu}^2 $ \cite{GAUGEFIX}.
 In this way one can define a {\em coarse grained} topological charge
 density by taking the sum on the right hand side of (\ref{eq07})
 in this ``maximally smooth'' gauge over an arbitrary domain of
 plaquettes.
 This gives information on the size and distribution of topological
 objects.

 The r\^ole of the gauge kinks can be important for quantities
 sensitive to the topological charge, or to the smoothness of the
 gauge field.
 Therefore it is instructive to study the model with some modified
 actions suppressing kinks.
 One way to push in the $\beta \to \infty$ limit all link
 angles to zero is to introduce a modified gauge field action like
\be \label{eq09}
S_4 = 16\beta_4 \sum_x \sum_{\mu=1, \nu=2}
\left[ 1 - cos(\Theta_{x\mu\nu}/4) \right] \ .
\ee
 The factor 16 in front of $\beta_4$ is introduced in order to have
 in the continuum limit of the action the same normalization as for
 $\beta$.

\begin{figure}
\vspace{5.2cm}
\includegraphics{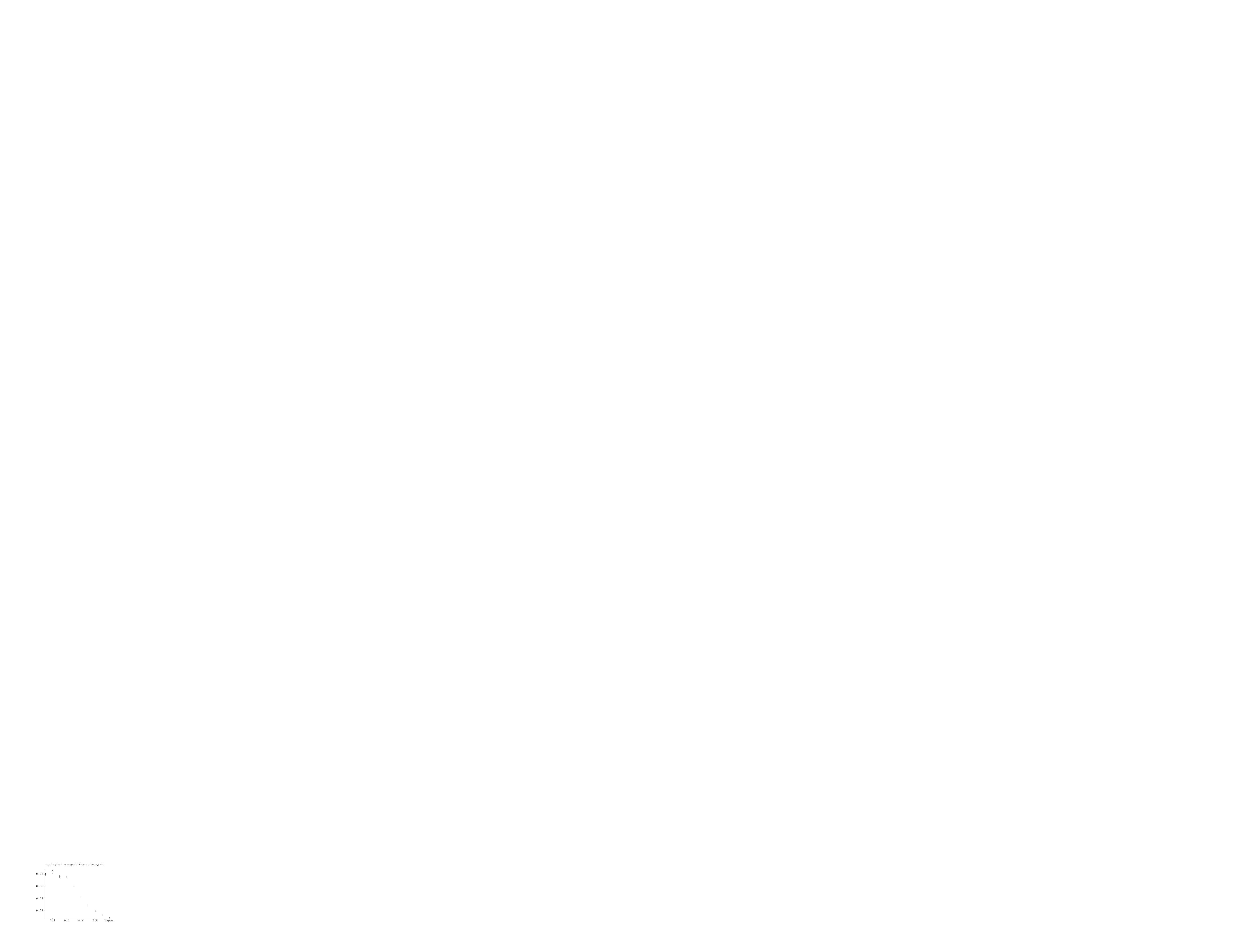}
\caption{ \label{figt4}
The topological susceptibility with the modified action $S_4$
as function of $\kappa$ at $\beta_4=2.0$ on $64^2$ lattice.
$\Omega\chi$ is plotted.}
\end{figure}

 Numerical simulations show that at $\beta_4=2.0$ the topological
 susceptibility is considerably suppressed by the action (\ref{eq09}):
 compare figs. \ref{figt2} and \ref{figt4}.
 At $\beta_4=8.0$ in a long run it turned out to be impossible
 to create even a single configuration with non-zero topological
 charge.

 An important question for future studies is to investigate the
 continuum limit of the anomaly equation (\ref{eq04}) and compare it
 to different definitions of the topological charge density.



\begin{thebibliography}{99}
%
\bibitem{THOOFT}
G. 't Hooft,
Phys. Rev. Lett. 37 (1976) 8; \\
Phys. Rev. D14 (1976) 3432.
%
\bibitem{BANKS}
T. Banks,
Phys. Lett. B272 (1991) 75.
%
\bibitem{TSUKPR}
I. Montvay,
Nucl. Phys. B (Proc. Suppl.) 26 (1992) 57.
%
\bibitem{PETCHER}
D.N. Petcher,
Nucl. Phys. B (Proc. Suppl.) 30 (1993) 50.
%
\bibitem{AMSTPR}
I. Montvay,
Nucl. Phys. B (Proc. Suppl.) 30 (1993) 621.
%
\bibitem{JOHOPR}
I. Montvay,
DESY preprint 93-134 (1993),
to appear in {\it Proc. of the 17th Johns Hopkins Workshop},
Budapest, July 1993.
%
\bibitem{BOUNDS}
G. M\"unster, contribution to Lattice '93;  \\
L. Lin, I. Montvay, G. M\"unster, M. Plagge, H. Wittig,
Phys. Lett. B317 (1993) 143.
%
\bibitem{BOCSHA}
A.I. Bochkarev, M.E. Shaposhnikov,
Mod. Phys. Lett. A2 (1987) 991.
%
\bibitem{STLOUIS}
I. Montvay, contribution to the 2nd IMACS Conference, St. Louis,
October, 1993.
%
\bibitem{LUSCH}
M. L\"uscher,
Comm. Math. Phys. 85 (1982) 39.
%
\bibitem{GKSW}
M. G\"ockeler, A.S. Kronfeld, G. Schierholz, U.-J. Wiese,
Nucl. Phys. B404 (1993) 839.
%
\bibitem{DEGTOU}
T.A. DeGrand, D. Toussaint,
Phys. Rev. D22 (1980) 2478.
%
\bibitem{GAUGEFIX}
I. Montvay, to be published.
%
\end{thebibliography}
\end{document}